\documentclass[draftcls,onecolumn,twoside,letter,12 pt ]{IEEEtran}
\usepackage{graphicx,subfigure}
\usepackage{caption}
\usepackage[cmex10]{amsmath}
\newcounter{subeqn} %
\makeatletter
\@addtoreset{subeqn}{equation}
\makeatother

\usepackage{algorithm}
\usepackage{algpseudocode}

\usepackage{epstopdf}
\usepackage{epsfig}
\usepackage{balance}
\usepackage{enumitem}
\usepackage{hyperref}
\usepackage[noadjust]{cite}
\usepackage{url}

\newcommand{\bs}{\boldsymbol}

\begin{document}
	\title{Blockchain-based Content Delivery Networks: Content Transparency Meets User Privacy}
	
	\author{\IEEEauthorblockN{Thang~X.~Vu, Symeon~Chatzinotas, and Bj\"orn~Ottersten}\\
	Interdisciplinary Centre for Security, Reliability and Trust (SnT) -- University of Luxembourg, Luxembourg\\
	Email: \{thang.vu, symeon.chatzinotas, bjorn.ottersten\}@uni.lu.}
	
	
	\providecommand{\keywords}[1]{\textbf{\textit{Index terms---}} #1}
	
	\date{}
	
	\maketitle
	\thispagestyle{plain}
	\begin{abstract}
		Blockchain is a merging technology for decentralized management and data security, which was first introduced as the core technology of cryptocurrency, e.g., Bitcoin. Since the first success in financial sector, blockchain has shown great potentials in various domains, e.g., internet of things and mobile networks. In this paper, we propose a novel blockchain-based architecture for content delivery networks (B-CDN), which exploits the advances of the blockchain technology to provide a decentralized and secure platform to connect content providers (CPs) with users. On one hand, the proposed B-CDN will leverage the registration and subscription of the users to different CPs, while guaranteeing the user privacy thanks to virtual identity provided by the blockchain network. On the other hand, the B-CDN creates a public immutable database of the requested contents (from all CPs), based on which each CP can better evaluate the user preference on its contents. The benefits of B-CDN are demonstrated via an edge-caching application, in which a feature-based caching algorithm is proposed for all CPs. The proposed caching algorithm is verified with the realistic Movielens dataset. A win-win relation between the CPs and users is observed, where the B-CDN improves user quality of experience and reduces cost of delivering content for the CPs.  
	\end{abstract}


	\section{Introduction}
	
During the past 40 years, the Internet has followed an extraordinary evolution and has become an integral part of the modern society. However, this evolution has kept momentum and there are constantly new services and contents distributed through this global communication network. According to Cisco's report~\cite{Cisco}, it is forecast that the mobile data traffic will grow $74$\% by $2021$. The main causes of this traffic growth are the vast availability of mobile handsets, e.g., smart phones, tablets, and notebooks, as well as the increasing growth of high-resolution video content on the Internet. Such availability of mobile devices has shifted the network traffic from traditional linear broadcasting services (TV channels) to streaming services, such as YouTube and NetFlix. Another factor that contributes to the traffic is the increasing video quality, i.e., 3D, 4K video, Virtual Reality etc., which eventually results in increased bandwidth requirements for both the core and access networks. From a telecom operator’s point of view, this increased video traffic will become a bottleneck and put excessive strain on current communication networks. These challenges have put pressure on both telecom operators and content providers to build efficient content delivery networks (CDNs), such as joint force between the Maldives operator with Google \cite{CDN}, and Orange with Akamai \cite{akamai}. The benefit of such collaboration is that the operators have access both to the physical infrastructure and the network services and this allows for a higher degree of cross-optimization and transmission efficiency. It becomes obvious that closer interaction between operators and content providers will be needed in order to optimize content delivery and overcome the projected bottleneck due to video traffic \cite{liv:edg:pro:cach:2014}. While a content provider tends to collaborate with the telecom operators, there is less joint effort among the content providers so far. This is because there lacks of an efficient and secure sharing platform connecting the content providers.

Blockchain is a merging technology for decentralized management and data security, which was first introduced as the core technology of cryptocurrency, e.g., Bitcoin \cite{Bitcoin}. Unlike centralized methods, there is no fixed controlling node in the blockchain network (BCN). When there is a new block to be added to the chain, a consensus mechanism will be implemented to vote a temporary controlling node. Several consensus mechanisms are used, e.g., proof-of-work (PoW), proof-of-stake (PoS), proof-of-capacity (PoC), which usually are the search for the answer of a difficult computational puzzle \cite{Bitcoin}. A blockchain node (or miner) who gets the first answer, which is called a \emph{nonce}, will broadcast the nonce to the network. Upon receiving the nonce, the blockchain nodes validate the new nonce with the existing chain. A block in the blockchain comprises a digest (a hash code) of the previous block. Therefore, any modification in a block will destroy the integrity of the chain. In short, the nature architecture of blockchain offers key features of security, temper-proof, and decentralization \cite{Bitcoin,SmartContract}.

Since the first success in the financial sector, blockchain has shown a great capability in wireless networks, with particular potentials in internet of things (IoT) \cite{Han,Cha2018,Novo2018,Sharma2018} and mobile networks \cite{Goka,Sharma2018a,Fan2018,Backman2017,Niyato:BC-cache}. The authors in \cite{Novo2018} propose a general architecture for blockchain-based IoT applications, which has desirable features for IoT services such as mobility, accessibility, scalability, lightweight, and transparency. The blockchain connected gateway architecture is proposed in \cite{Cha2018} for bluetooth low energy applications, which creates a secure platform between the user and IoT devices' owner or manager via smart contract. 
The authors in \cite{Han} evaluate the performance of the Byzantine consensus under IoT applications via a private BCN. In this work, the Byzantine consensus is implemented on the Hyperledger Fabric (HLF) platform to measure the throughput and mean latency as a function of IoT requests. A secure and distributed management for mobile ad-hoc networks is proposed in \cite{Goka}, which manages reputation and reward of a node securely  via the BCN. In \cite{Sharma2018a}, a blockchain-based handover management framework is proposed for fog radio access networks. Focusing on the future factory, the authors in \cite{Backman2017} propose a network slicing broker based on blockchain to promote the interaction between multi-tenants and infrastructure providers via a shared slicing ledger.
The authors in \cite{Fan2018} propose a privacy-preserving and data sharing scheme for content centric networks based on public BCN. An autonomous content caching
market is proposed in \cite{Niyato:BC-cache} in which the negotiation between content providers (CPs) and cache helpers is modelled as a Chinese restaurant game. 

In this paper, we propose a novel architecture for content delivery networks based on blockchain (B-CDN) which creates a decentralized and secure management platform for different CPs and users. Compared with the conventional CDN architecture, the proposed B-CDN offers various benefits to both CPs and users. Firstly, the B-CDN will leverage the user inscription to different CPs while guaranteeing the user privacy via virtual identity representation. Secondly, the CPs can considerably minimize expense in managing their customers' subscription since these tasks are managed by the BCN. Last but not least, a public and immutable database is created by the proposed B-CDN, from which the CPs can exploit to maximize their service efficiency. The benefits of B-CDN are demonstrated via an edge-caching application, in which the CPs exploit the public database available on the BCN to develop their caching algorithm. A win-win relation between the content providers and customers is observed, where the B-CDN improves user quality of experience and reduce cost of delivering content for all CPs. In particular, a feature-based caching algorithm is proposed to improve cache hit ratio (CHR) and reduce delivery time for all CPs. Finally, the effectiveness of the proposed caching algorithm is validated via numerical results based on the realistic Movielens dataset.

The rest of this paper is organised as follows. 
Section~\ref{sec:sysmodel} presents the proposed blockchain-based CDN model. Section~\ref{sec:caching} develops an edge caching algorithm based on the proposed B-CDN. Numerical results are shown in Section~\ref{sec:result}. Finally, Section~\ref{sec:conclusion} provides conclusions and discussions. 
	
%

\begin{figure*}
	\centering
	\includegraphics[width=0.9\textwidth]{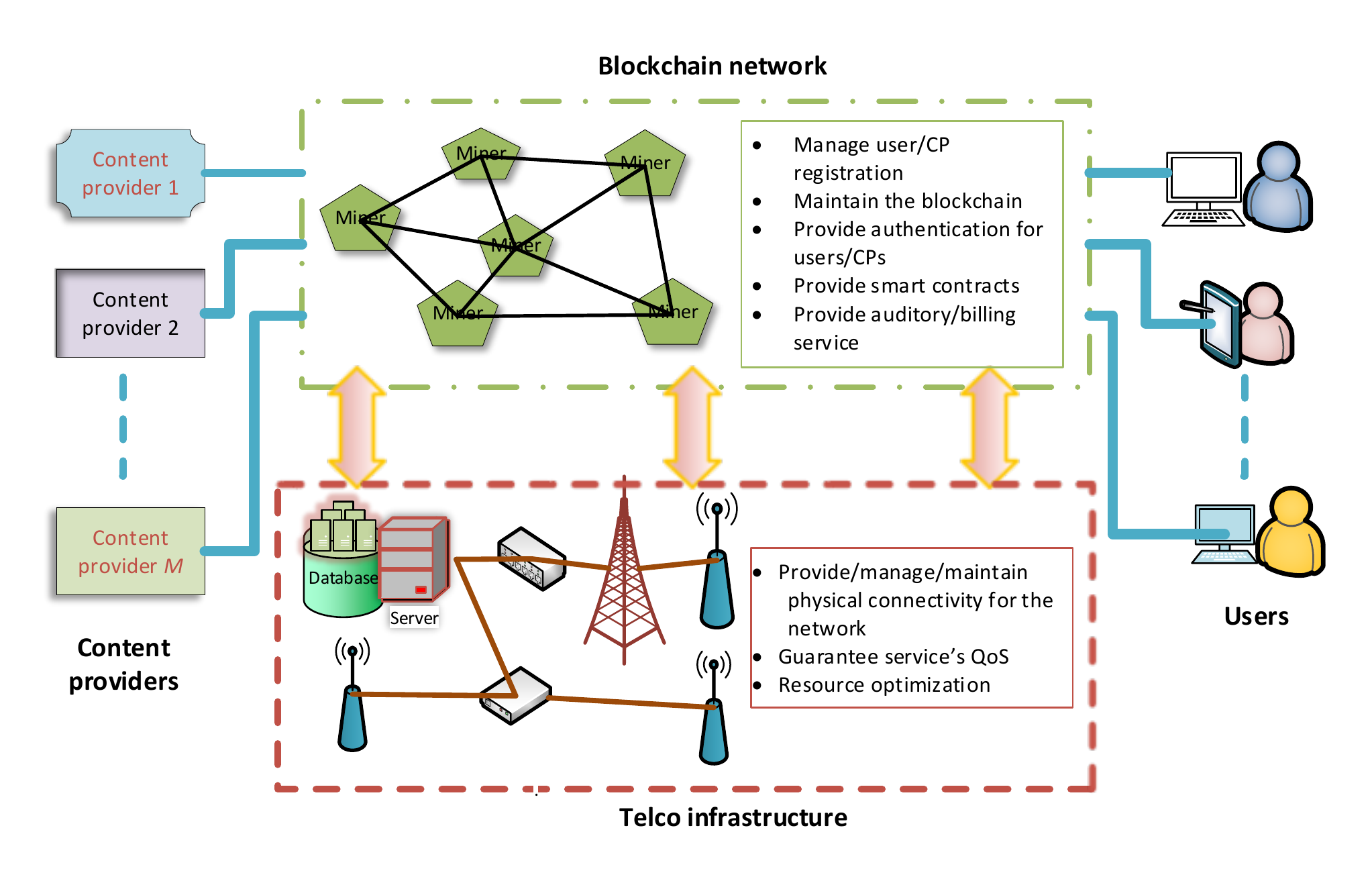}
	\caption{System architecture of the proposed Blockchain-based Content Delivery Networks.} \label{fig:systemmodel}
	\vspace{-0.3cm}
\end{figure*}

\section{Proposed Blockchain-based Content Delivery Networks}\label{sec:sysmodel}

In conventional CDN architectures, the users have to subscribe multiple times when they want to use different services from different CPs since the CPs usually do not cooperate due to conflict of interest among the CPs and user privacy constraints. In this section, we propose a novel B-CDN architecture based on the blockchain technology, which provides content transparency and assures user privacy to non-registered CPs. The high-level block diagram of the proposed B-CDN is depicted in Fig.~\ref{fig:systemmodel}. There are four stakeholders in the proposed B-CDN: i) telecommunication provider (Telco), ii) content providers, iii) customers (users), and iv) blockchain network.

\begin{itemize}
	\item \emph{The Telco} acts as the foundational infrastructure that guarantees physical connections among partners in the network. In particular, the Telco will: a) provide/manage/maintain physical connectivity in the network, b) guarantee the target quality of service, and c) optimize the resource allocation.
	\item \emph{Content providers (CPs)} sell contents and services to prospective customers or users, e.g., movies and entertainment programs. A CP can be an independent party, e.g., NetFlix, or an alliance with the Telco, e.g., Orange and Akamai \cite{akamai}. 
	\item \emph{Users} are the customers who buy contents and services provided by the CPs.
	\item \emph{Blockchain network (BCN)} acts as the core management entity that provides a decentralized and secure data management for the B-CDN. The BCN is a peer-to-peer network of blockchain nodes (or miners) which perform validating and adding new blocks to the chain of "blocks". In order to improve the resource efficiency, a private blockchain is employed and practical Byzantine fault tolerance (PBFT) \cite{PBFT} is used as the consensus method. The BCN will be responsible for: a) managing CPs and users registration, b) providing smart contracts and maintaining the blockchain, c) providing authentication for both users and CPs, and d) providing accounting and auditory service. The owner  can be the Telco or an independent partner.
\end{itemize}

\subsection{Operations in the proposed B-CDN}
The B-CDN is operated via two main activities: \emph{authority registration} and \emph{service subscription}. When a CP or user joints the B-CDN, it first has to register with the BCN. Successful registrations (of CPs or users) are done by transaction between the BCN and the registered entity (CP or user), which will be validated and added to the blockchain. Service subscription occurs between a registered user with one or several registered CPs, which is implemented via smart contracts \cite{SmartContract}.

\subsubsection{CP registration}
A CP is represented via a pairs of keys: a private key $K^c_{pri}$ and a public key $K^c_{pub}$. To register with the BCN, the CP sends a registration request to a blockchain node (miner) containing its public key, address and other public information. The blockchain node then sends these information to the BCN as a transaction that is then added to the blockchain. 

\subsubsection{User registration}
A user is described by a pairs of a private key  $K^u_{pri}$ and a public key $K^u_{pub}$. In order to protect the user privacy, virtual identity  \cite{Lee2018} is employed to represent the user in B-CDN. First, the user uses its public key to generate the virtual identity $\mathtt{VID}^u$, which will be used to register with the BCN. The virtual identity can be generated from the public key as the hash value of its public key, e.g.,  $\mathtt{VID}^u = \mathrm{hash}(K^u_{pub})$. The user then sends its public key and virtual identity to the BCN for registration. Next, a blockchain node, e.g., the base station serving the user, generates a signature on the user's public key and virtual identity $\mathtt{Sig}_{K^b_{pri}}(K^u_{pub}, \mathtt{VID}^u)$, where $K^b_{pri}$ is the private key of the blockchain node. The blockchain node then sends the user's public key, virtual identity and the generated signature to the BCN as a transaction. After registration, all the blockchain nodes in the BCN store the user's public key and virtual identity. 

It is worth noting that only the virtual identity is available in the public ledger of the BCN, hence the user privacy is guaranteed. Only the CPs to whom the user subscribes can access to the user's public key, as details in the next subsection. 

\begin{figure}
	\centering
	\includegraphics[width=0.6\columnwidth]{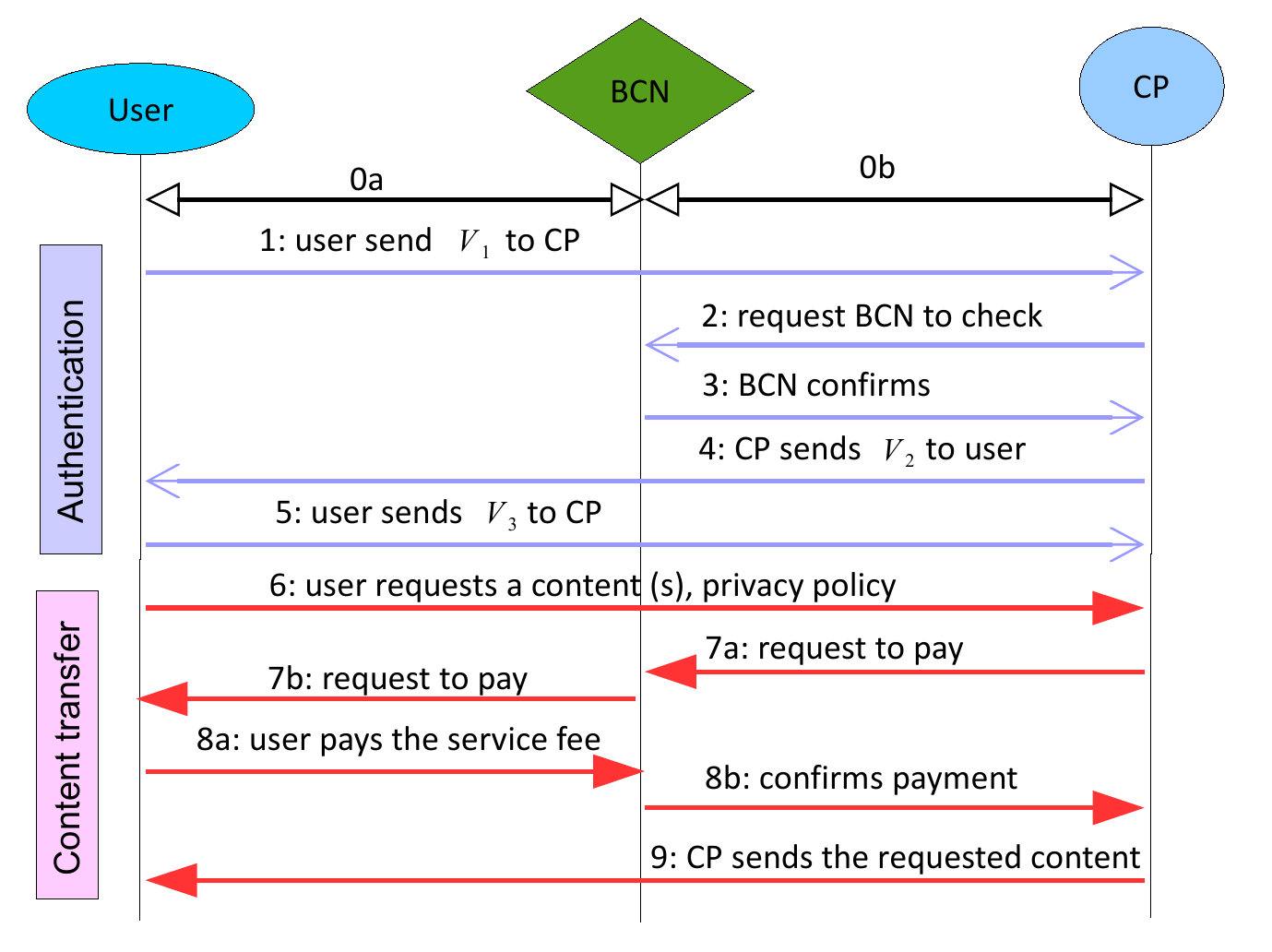}
	\caption{Information flows between the BCN, CP and user for a service registration and smart contract.} \label{fig:workflow}
	\vspace{-0.3cm}
\end{figure}
\subsubsection{Service subscription and smart contract}
To use a service provided by the CP, e.g., download a TV series, the user has to subscribe to the CP. We adopt the method in \cite{Lee2018} to manage the service subscription and authentication via 5 steps, as depicted in Fig.~\ref{fig:workflow}. First, the user sends a message $V_1 = (\mathtt{VID}^u, \mathtt{non}, \mathtt{Sig}_{K^u_{pri}}(\mathtt{VID}^u, \mathtt{non}))$ to the CP, which consists of the user's virtual identity $\mathtt{VID}^u$, a nonce $\mathtt{non}$, and the signature $\mathtt{Sig}_{K^u_{pri}}(\mathtt{VID}^u, \mathtt{non})$ using the user's private key. Upon receiving the request, the CP asks the BCN for validating $\mathtt{VID}^u$. If the user has already registered to the BCN, i.e., $\mathtt{VID}^u$ is valid, then the CP can obtain user's public key, $K^u_{pub}$, and message $V_1$. In the next step, the CP includes its public key $K^c_{pub}$ into his signature signed using the obtained public key from the user. This signature is then sent together with the user's virtual identity and nonce $\mathtt{non}+1$ to the user in message $V_2 = (\mathtt{VID}^u, \mathtt{non}+1, \mathtt{Enc}_{K^u_{pub}}(\mathtt{VID}^u, \mathtt{non}+1, K^c_{pub}))$. After successfully receiving the message, the user decrypts with its private key (public key is generated from the private key) to get the CP's public key $K^c_{pub}$ and validates $\mathtt{non}+1$. Then the user sends a message $V_3 = (\mathtt{VID}^u,\mathtt{non}+2,\mathtt{Enc}_{K^c_{pub}}(\mathtt{VID}^u,\mathtt{non}+2))$. As the CP receives $V_3$, it decrypts with $K^c_{pri}$ and validates the message with its private key and the authentication is completed. 

When the authentication is completed, the user can request a service provided by the CP via a smart contract, which is a mutual agreement between the CP and user on the requested service and a fee for using the service. First, the user sends a \emph{service request} message to the CP. This message includes the virtual identity of the user together with the service name. The service can be a flat-rate plan or on demand contents. Upon receiving the request, the CP verifies the user virtual identity and the requested service. If the verification succeeds, the CP sends a request-to-pay message to the user via the BCN. When the payment is confirmed by the BCN, the CP will send the requested service to the user and completes the smart contract. Finally, the BCN inserts a timestamp and then binds the contract (together with other smart contracts from other CPs) to the blockchain. 
\subsection{Benefits of the proposed B-CDN}
The proposed B-CDN provides benefits to not only CPs but also other stakeholders in the network. 
\subsubsection{Benefits to CPs}
The proposed B-CDN positively provides multiple benefits to the CPs. Firstly, by tracking the transaction history in the ledger available in the BCN's database, a CP can establish the global information on the content popularity and user preference, based on which the CP minimizes operating expense and improves user quality of experience. Secondly, since user registration and management are done by the BCN, the CPs can save their own resource in creating and maintaining an in-house identity and authentication management infrastructure. 

\subsubsection{Benefits to users}
The first benefit to users is the privacy, since each user is represented in the BCN by its virtual identity. Despite using virtual identity, the BCN guarantees for the true mapping from virtual identity to the physical address. More importantly, with virtual identity, the user does not need to register multiple times when it uses services from different CPs. In addition, by allowing some users' common information available in BCN, the CPs get to know more users' preference, hence can serve them in an efficient manner. 

\subsubsection{Benefits to BCN}
As the backbone management platform, the BCN provides a efficient way to connect users and CPs. By providing the connecting services, the BCN can receive incentives via transaction fees or service fees. These fees can be charged from CPs as parts of users' registration/membership fee. 

\subsubsection{Benefits to Telco}
As the infrastructure provider, the Telco gets benefit from other stakeholders, e.g., users, CPs, and BCN (if not owned by the Telco).

\section{Blockchain-based Edge Caching in B-CDN}\label{sec:caching}
In this section, we demonstrate the benefit of the proposed B-CDN via an edge-caching application. In order to improve the quality of experience and reduce transmission cost, the CPs install distributed caches at edge nodes, e.g., base stations or dedicated cache helpers, and develop a caching algorithm to prefetch some contents in the caches. If the requested file is available at the edge node's cache, it can be served immediately without being sent from the core network \cite{Vu18twc,Vu18wcl}. Therefore, a proper caching algorithm not only shortens content delivery time but also reduces backhaul traffic. Obviously, if the (temporary) content popularity is known in advance, each CP can simply put the most popular contents to the edge nodes' cache until full. In practice, it is unfortunately very challenging to obtain these information because the network topology and user preference are highly dynamic. Therefore, a favour caching algorithm should not only correctly match with temporary content popularity but also be able to predict future user preference. 

Assume that $M$ content providers, denoted by ${CP}_m, m = 1, \dots, M$, employ the B-CDN platform to serve a set of $K$ users, denoted by $\mathcal{U} = \{1, ..., K\}$. The ${CP}_m$ owns a library of $N_m$ contents, denoted by $\mathcal{N}^m = \{c^m_1, \dots, c^m_n, \dots, c^m_{N_m}\}, \forall m$. The global library from all CPs is denoted by $\mathcal{N} = \cup\mathcal{N}^m$, whose cardinality is $N = |\mathcal{N}| \leq \sum_{m=1}^M N_m$, since different CPs may have some common contents. It is worth noting that one CP may serve only a subset of $\mathcal{U}$. Our goal is to estimate the temporary content popularity for each CP, from which proper caching algorithms can be developed. Our proposed caching algorithm differs from \cite{Niyato:BC-cache} in three aspects: i) we consider the realistic scenario that different CPs own different content libraries, while \cite{Niyato:BC-cache} assumes all CPs share a common library; ii) we develop the caching algorithm based on content's feature similarity model, while \cite{Niyato:BC-cache} simply counts the requested contents; iii) we validate the proposed caching algorithm via the realistic Movielens requests, while \cite{Niyato:BC-cache} is based on the Zipf-like content popularity model. 

\subsection{Feature-based model and feature popularity}
In the feature-based model, each content is represented via a list of features. A feature can be any verbal and numeric information about the content, e.g., type and production year of a movie, which is usually available in the content's metadata. Denote $\bs{F}_n = \{F_{n,1}, \dots, F_{n,l}, \dots, F_{n,L}\}$ is the $L$-vector of features of the $n$-th content, where $F_{n,l} = 1$ if content $n$ has the $l$-th feature and $F_{n,l} = 0$ otherwise. 

In order to predict the feature popularity, each CP first requests to have access to all transactions history available in the BCN, i.e., the public ledger. If the CP's address and signature are well verified, the BCN sends the requested list to the CP. This list contains the metadata of the contents requested by all users in the network. Next, the CP determines the content features which attract users the most via feature popularity estimation, which is done in 2 steps: i) compute the total number of requests for each feature, and ii) sort the features' counts in a decreasing order. The estimated feature popularity can be seen as the global popularity since the estimation is based on the users' requests from all CPs. Hence, it captures the user preference better than the estimation based only on current registered users of individual CP. It is worth noting that, despite knowing all transactions history in the BCN, the user privacy is not invaded since only virtual user identities are recorded in the ledger. 

\subsection{Content popularity estimation and caching algorithm}\label{sec:caching pop}
From the estimated feature popularity, CP  $m$ can estimate the user preference on its own contents in $\mathcal{N}^m$, by calculating the correlation between the CP's contents and the estimated feature popularity. In this work, we employ cosine similarity \cite{Lee:1999:vectorsimilarity} to measure the correlation between a content and the global feature popularity. Let $p_n$ denote the correlation between content $n$'s feature vector, $\bs{F}_n$, and the feature popularity, $\bs{q}_\star$, which is calculated as
\begin{align}	
	p_{n} = \frac{{\bs{F}_n}^T \bs{q}_\star }{{\parallel \bs{F}_n\parallel}  {\parallel \bs{q}_\star\parallel}},~\forall n  \in \mathcal{N}^m,
\end{align} 
where $(.)^T$ and $\parallel.\parallel$ denote the transpose and the $l_2$ norm. 
 
Denote $\bs{p}^m = \{p_n\}_{\forall n\in \mathcal{N}^m}$ as the correlation vector of all contents from CP $m$. The CP $m$ then sorts $\bs{p}^m$ in the decreasing order to obtain $\bs{p}^m_\star$, the sorted content popularity at CP $m$. Finally, the first $Z$ contents in $\bs{p}^m_\star$ are prefetched in the CP $m$'s cache, where $Z$ is the cache size. The pseudo-code of the proposed caching algorithm is given in Algorithm~\ref{algo:1}.

\begin{algorithm}
	\caption{Blockchain-based edge caching algorithm}\label{algo:1}
	\textsc{A - Feature popularity extraction:}
	\begin{algorithmic}[1]
		\State  Request transactions history (ledger) from the BCN
		\State $\bs{q} \gets \mathit{zeros}(1,L)$
		\State \textbf{for} {every content $n$ in the ledger} \textbf{do}: 
		$\bs{q} = \bs{q} + \mathbf{F}_n$		
		\State $\bs{q} = \frac{\bs{q}}{|\bs{q}|}$
		\State $\bs{q}_\star \gets$ sort($\bs{q}, `descend$')	
	\end{algorithmic}
\textsc{B - Content popularity estimation and caching for the $m$-th CP:}
\begin{algorithmic}[1]
	\State  $\bs{p}^m \gets zeros(1, N_m)$
	\State \textbf{for} every content $n \in \mathcal{N}^m$ \textbf{do}: 
	 	$\bs{p}^m[n] = \frac{{\bs{F}_n}^T \bs{q}_\star }{\sqrt{{\parallel \bs{F}_n\parallel}^2} \sqrt{ {\parallel \bs{q}_\star\parallel}^2}}$
	\State $\bs{p}^m_\star \gets \text{sort}(\bs{p}^m, `\textit{descend'})$
	\State Prefetch $Z$ first contents in $\bs{p}^m_\star$ in the cache	
\end{algorithmic}
\end{algorithm}

%


\section{Numerical results}\label{sec:result}
In this section, we demonstrate the advantage of the proposed edge caching algorithm. In the considered system, three CPs employing the B-CDN to serve the users. The contents and user requests are taken from the realistic Movielens dataset \cite{movielens20M}.  

\subsection{Movielens 20M dataset}
The contents and user requests are obtained from the Movielens 20M dataset, which comprises 20 million ratings applied to 27,000 movies by 138,000 users. In this dataset, each user is assigned a unique $\mathtt{userId}$. Similarly, each movie is represented by a unique $\mathtt{movieId}$. All the ratings are listed in file rating.csv, in which each line shows a specific user has rated a movie and the time the rating was recorded. Unlike \cite{liv:edg:pro:cach:2014}, which develops a recommendation system by guessing the preference of the users on particular movie based on matrix completion, we do not take into account the rating score and consider each rating as one user request. We extract the ratings during one year from March, 2014 to March, 2015. A disjoint random selection of 1000 movies is assigned to each CP. It is noted that there is no common movie between the two CPs. The movies are featured by $L = 18$ genres, listed in file movies.csv. The 18 genres are: action, adventure, animation, children's, comedy, crime, documentary, drama, fantasy, film-noir, horror, musical, mystery, romance, sci-fi, thriller, war, and western.

\subsection{Performance evaluation}
We consider a scenario that CP1 is already running its services on the B-CDN and serves parts of the users. Two content providers CP2 and CP3 have just joined the B-CDN and lunched their services. All the CPs develop their caching algorithm according to Algorithm~\ref{algo:1}. We compare the proposed B-CDN based caching algorithm with the one based on the conventional CDN architecture, in which one CP can only access to requests history from its own contents and has no knowledge about the requested contents from other CPs. Because CP2 and CP3 have just lunched their service, there is no request history on their contents. As a result, a random caching policy is employed by CP2 and CP3 in the conventional architecture. In the proposed B-CDN architecture, however, by reading all transactions in the ledger provided by the BCN, both CP2 and CP3 can access to the request history of CP1, from which they evaluate user preference on their contents. On the other hand, since CP1 has the knowledge on the request history from its own users, the caching algorithm for CP1 is similar in both B-CDN and conventional CDN architectures.

\begin{figure}
	\centering
	\includegraphics[width=0.6\columnwidth]{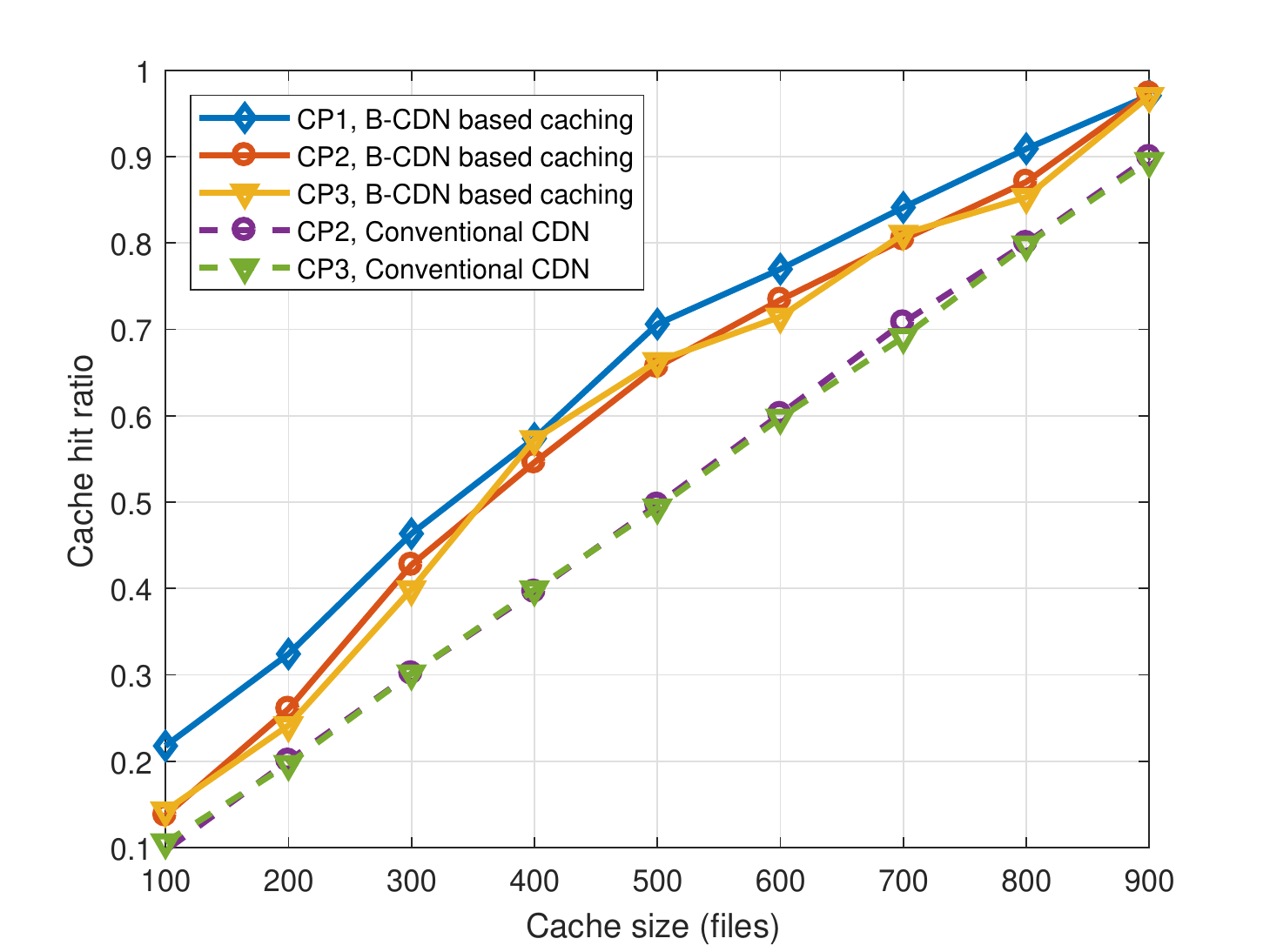}
	\caption{Cache hit ratio performance of the proposed B-CDN-based caching algorithm compared with the conventional CDN architecture. The performance of CP1 is the same in both blockchain-based and conventional architectures since the caching algorithm is based on its own request history.} \label{fig:CHR}
\end{figure}

Fig.~\ref{fig:CHR} presents the CHR of the proposed caching algorithm as the function of the cache size. It is not surprised that CP1 achieves the highest CHR because the estimated content popularity is based on the CP1's request history. However, CP2 and CP3 under the proposed B-CDN, whose caching algorithm is based on the demand records of CP1, also achieve a close performance as CP1. Compared with the conventional CDN architecture, the B-CDN increase the CHR for both CP2 and CP3 by $30\%$ at the cache size of 500 (files). This gain clearly shows the benefits of the proposed B-CDN architecture, which improves the average CHRs of all CPs in the network. 

\begin{figure}
	\centering
	\includegraphics[width=0.6\columnwidth]{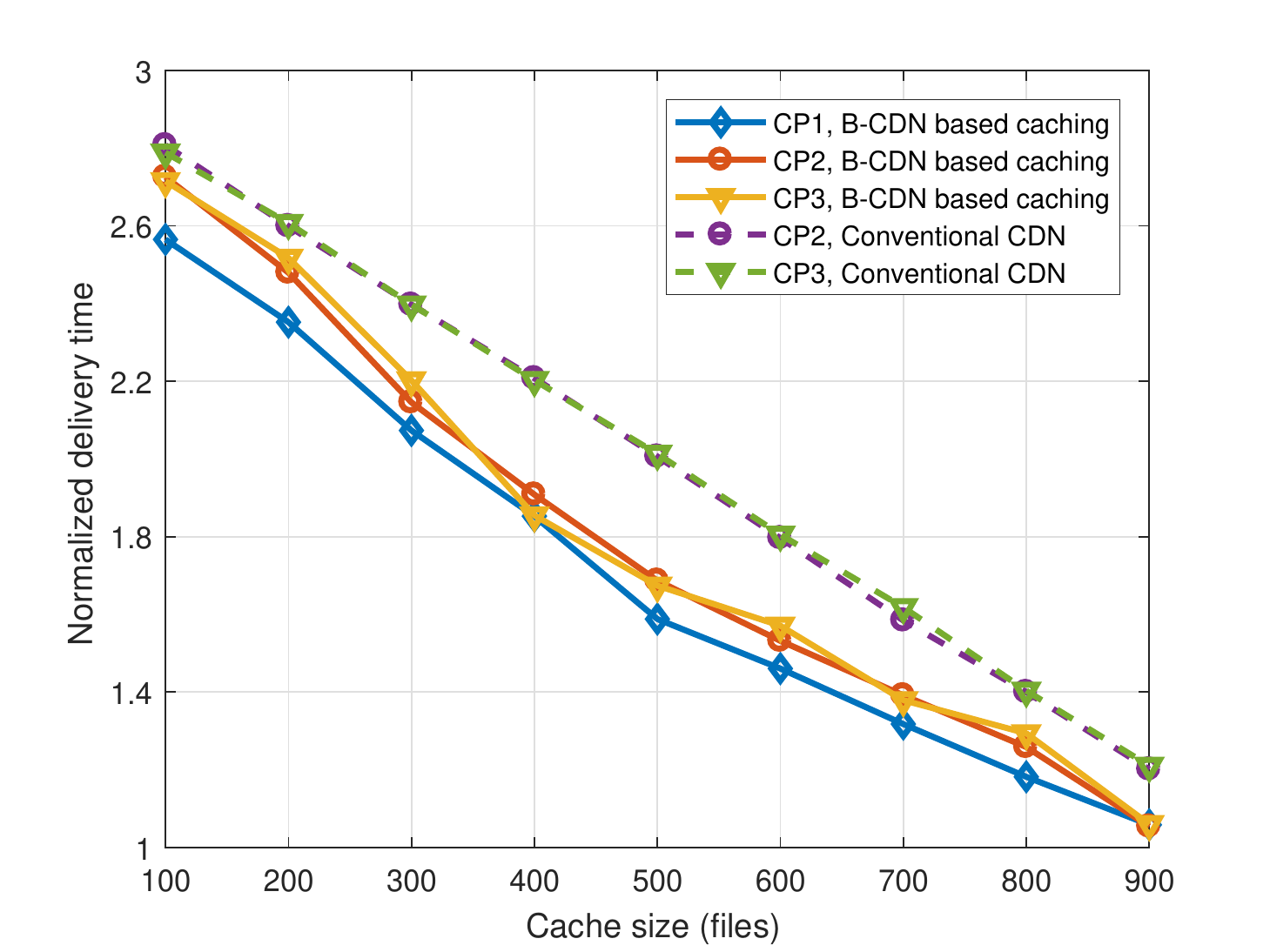}
	\caption{Normalized delivery time performance of the caching algorithms proposed for the B-CDN compared with the conventional CDN architecture.} \label{fig:DeliveryTime}
\end{figure}

Fig.~\ref{fig:DeliveryTime}) shows quality of experience (QoE) improvements brought by the proposed B-CDN architecture via the normalized delivery time metric. The delivery time to serve a request for content $n$ is calculated as:
\begin{align*}
\tau_{tot} = \left\{  
\begin{array}{ll}
\tau_{AC},&~\text{if file $n$ is cached}\\
\tau_{BH} + \tau_{AC},&~\text{if file $n$ is not cached} 
\end{array},
\right. 
\end{align*}
where $\tau_{AC}$ is the delivery time from the edge node to the user, and $\tau_{BH}$ is the delivery time from the network core to the edge node. The normalized delivery time is computed as $\frac{\tau_{tot}}{\tau_{AC}}$. In general, the normalized delivery time is a decreasing function of the cache size. A similar conclusion is drawn that the caching algorithm based on B-CDN significantly reduces the delivery time for both CP2 and CP3 compared with the caching algorithm based on the conventional architecture, hence improves the user quality of experience. At the cache size of 500 (files), the B-CDN based caching algorithm reduces the delivery time in CP2 and CP3 by 15$\%$ compared with the conventional CDN architecture. 
 	\balance
\section{Conclusion and Discussion}\label{sec:conclusion}
We have proposed a novel blockchain-based content delivery network architecture, which potentially provides benefit to all partners while guaranteeing the user privacy. The advantage of the proposed architecture is demonstrated via an edge caching application, in which a feature-based caching algorithm is employed to maximize the cache hit ratio for all content providers. The effectiveness of the proposed caching algorithm is verified by the improvements in both cache hit ratio and quality of experience based on the realistic Movielens dataset. 

The proposed caching algorithm in Sec.~\ref{sec:result} relies only on one feature \emph{genres} of the movies in the Movielens dataset. In practice, the performance of the proposed caching algorithm can be improved by considering additional features, e.g., producer, thumbnail, actors etc. Using more features enables to construct a more accurate correlation model between the contents from different CPs, which eventually improves the estimation of content popularity. Furthermore, the caching algorithm might be further improved by considering user information together with the content features. In this way, each CP may target a specific group of users based on its available content. We note that since only virtual identities are shown in the blockchain network, the privacy of users is secured. 

\section*{Acknowledgement}
This work is supported by the Luxembourg FNR CORE program under project ProCAST (R-AGR-3415-10).

\balance
	

\end{document}